\documentclass[manuscript]{aastex}

\shorttitle{An Atlas of STIS-HST Spectra of Seyfert Galaxies}
\shortauthors{P.F.Spinelli et al.}
\begin{document}

\title{An Atlas of STIS-HST spectra of Seyfert Galaxies}

\author{P. F. Spinelli, T. Storchi-Bergmann, C. H. Brandt}
\affil{{\it Instituto de F\'{i}sica, UFRGS, Porto Alegre, RS, Brazil}}
\email{patricia.spinelli@ufrgs.br}
\and
\author {D. Calzetti}
\affil{{\it Space Telescope Science Institute, Baltimore, MD 21218}}

\begin{abstract}
We present a compilation of spectra of 101 Seyfert galaxies
obtained with the Space Telescope Imaging Spectrograph (HST-STIS),
covering the UV and/or optical spectral range. 
Information on all the available spectra have been collected in a 
{\it Mastertable}, which is a very useful tool for anyone interested in a quick glance 
at the existent STIS spectra for Seyfert galaxies in the HST archive, and it can be recovered 
electronically at the URL address {\bf www.if.ufrgs.br/\~{}pat/atlas.htm}. 
Nuclear spectra of the galaxies have been
extracted in windows of 0\farcs2 for an optimized sampling
(as this is the slit width in most cases),
and combined in order to improve the signal-to-noise ratio 
and provide the widest possible wavelength coverage. 
These combined spectra are also available electronically. 
\end{abstract}

\section{Introduction}

Spectra obtained with the Space Telescope Imaging Spectrograph
(HST-STIS) provide unique information on the spectral energy distribution 
(SED) of active galactic nuclei (AGN), in two aspects:
the coverage of the ultraviolet spectral range, which is not observable from the ground, and the high angular resolution which enhances the 
contrast between the nuclear continuum and that of the stars 
of the host galaxies. Now that STIS has ceased to work,
it is timely to compile the data accumulated by 
observations with this instrument in an Atlas. In
the present work we provide such compilation for
101 Seyfert galaxies. 

We have used the spectra to construct nuclear SED's
of Seyfert galaxies obtained from extractions at an optimized sampling,
corresponding to an aperture 0\farcs2 $\times$ 0\farcs2 , 
as 0\farcs2 is the width of the slit in most 
observations. These combined nuclear spectra are available electronically, and
can be used for a number of studies. 
The small extraction window allows us to better isolate the nuclear SED, minimizing the contamination by the bulge of the host galaxies.
These spectra can be compared with data obtained through large apertures using ground based telescopes in order to evaluate the contribution of the of host
galaxies, particularly useful when studying samples of
distant AGNs. These spectra can also be used to investigate
the contribution of other sources very close to the nucleus, such as
starbursts \citep{sb05,gd04}.  

Although the HST archive provides one-dimensional spectra, which are
identified by the terminations {\it \_x1d} and {\it \_sx1}, our Atlas has 
at least three advantages: 

(1) The {\it \_x1d} and {\it \_sx1} are obtained with a extraction window 
of 11 pixels for the UV corresponding to 0\farcs27, and 7 pixels for 
the optical -- corresponding to 0\farcs35. Therefore, the extraction windows 
in the UV an optical are different and do not make optimal use the 
angular resolution provided by HST. Our extraction window is chosen 
to have the same angular extent of the slit width, 0\farcs2 
in all wavelength ranges, providing spectra with better angular 
resolution.  For AGN, a smaller extraction window increases the 
contrast between the active nucleus and the host galaxy.

(2) In many cases the HST pipeline does not perform averages of spectra. 
This is the case of the {\it \_x1d} spectra which are very noisy.

(3) The pipeline also does not "glue" the different spectral segments 
together. In the Atlas we have done this after eliminating the noisy borders of each spectral segment.

Our Atlas thus provides 
better signal-to-noise ratio nuclear spectra with the widest available 
spectral coverage, with the different spectral ranges already combined 
and edited to eliminate the noise usually present at the initial and 
final wavelengths of each segment.

In the process of constructing the Atlas, we have compiled relevant information on all the available spectra we have been collected in
in a {\it Mastertable}. It contains for example, initial and final
wavelengths of the different spectra segments, exposure times, gratings and slit width. This {\it Mastertable} is
by itself a very useful tool for anyone interested in a quick glance 
at the available STIS spectra for Seyfert galaxies in the HST archive
and can be recovered electronically as the spectra.

This paper is organized as follow: section 2 describes our sample selection. Section 3 presents the {\it Mastertable} and describes the information contained in it. The extraction of the spectra
is described in section 4 and their combination is explained in section 5. The results and some potencial applications are discussed in section 6.

\section{Sample and Data}

The sample was initially selected as all Seyfert galaxies listed in the
catalog of \citet{ver96} with redshift z$\le$0.03, which had STIS spectra
available in the archive.  We have later tried to incorporate the remaining
Seyfert galaxies (z$\geq$0.03). Misclassification,
however, may have prevented a comprehensive inclusion of all Seyfert galaxies
in the HST archive. Thus our sample comprises most galaxies (101 in the total) classified as Seyfert  with available  STIS spectra in the HST archive until September 2004. Although the most valuable wavelength range
is the UV because it is not accessible from the ground, we have
included in the Atlas also those cases in which only optical spectra
were available. The sample galaxies are listed in Table 1, which contains information on the positions, Hubble type, activity type, redshift and references to previous works in which the spectra have been used. 
The seventh column of Table 1 gives the spectral coverage (in the observed frame) of the resulting nuclear spectrum after the individual extractions and combination of the different spectral segments. 

\section{The Mastertable}

Relevant information about all the two-dimensional (hereafter 2-D) 
spectra collected is summarized in a {\it Mastertable}, 
available electronically in the URL address {\bf www.if.ufrgs.br/\~{}pat/atlas.htm}.
The columns of the table contain the following information:
(1) the name of the galaxy; (2) the identification of all 
available STIS spectra for this galaxy in the HST archive,
one per line; (3) the grating used in each observation; (4) the slit width of each observation;  (5) the central wavelength (in the observed frame); (6) the initial wavelength (in the observed frame); (7) the final wavelength (in the observed frame); (8) the spectral resolution; (9) the slit orientation; (10) the exposure time. In column (11) we list the identification of the extracted spectrum from each segment, which will be useful in a few cases in which we could not combine the spectra of the same galaxy (for example, because they were obtained in different slit positions) and we then provide the individual extracted spectra without combining them. These spectra are identified  according to following convention: 
compact name of the galaxy followed by an arbitrary ordering number 
and the slit orientation. For example, n3516-13.97  means the 13th spectrum of 
the galaxy NGC\,3516 which was obtained at slit orientation of 97 degrees. 
Finally, in column (12) we list the platescale of the observations.
In Table 2 we present a printout of a few selected lines of the {\it Mastertable} (which has 1001 lines), for illustrative purposes.

\section{Extraction of the spectra}

The nuclear spectra were obtained from 2-D reduced STIS spectra,
which have been rectified, wavelength and flux calibrated, and are 
identified in the HST archive by the suffixes {\it \_x2d} and {\it \_sx2}. The latter are summed {\it \_x2d} spectra (when the observations were performed in the {\it cr-split} or {\it repeatobs modes}).
When both  {\it \_x2d} and {\it \_sx2} spectra were available, we used the latter.

One-dimensional (hereafter 1-D) spectra were extracted from the 2-D spectra
in windows of 0\farcs2\
from a long-slit spectrum usually obtained through a slit width 
0\farcs2 and covering 52\,$^{\prime\prime}$ in the sky. 
The {\it IRAF} task  {\it apall} was used to perform the extractions. We performed the
sky subtraction by fitting a straight line to regions along the slit with no 
(or negligible) galaxy contribution.  For each galaxy, different sky windows were defined, 
in order to avoid including contribution from the galaxy. The sky level was always
negligible, except in the Lyman alpha Geocoronal emission line.

Although the redshift range for the sample is 0$\le$z$\le$0.37, only for 15\% of the galaxies $z\ge$0.06, such that the 0\farcs2 aperture corresponds at the galaxies to more than 200\,pc. For 60\% of the sample, 0\farcs2 corresponds to $<$ 60\,pc at the galaxies,  while for 30\% of the sample it corresponds to $<$ 20\,pc. We are thus sampling a very small region around the active nucleus, providing the best possible contrast between the AGN and galaxy bulge.

We extracted only nuclear spectra which we 
identified as being centered at the peak of the continuum flux along the slit.
This was done by inspecting the spatial light distribution in a spectral region 
devoid of emission or strong absorption lines (the continuum) 
and centering the extraction window at the peak of the continuum flux 
distribution. In a few cases the 2-D spectra contained only emission-lines,
with no continuum. In these cases, for which we could not identify a continuum 
source we did not extract the spectra, but the information on the available 2-D spectra are still listed in the {\it Mastertable}, with a cautionary note 
explaining why the spectra have not been extracted.

In the cases for which there were more than one continuum source we extracted the brightest one. Although we cannot be absolutely sure for all cases, Seyfert galaxies are usually the brightest object. There are two exceptions, for which we did not extract the spectra because the two sources were equally bright. These two cases have also been identified in the {\it Mastertable} with a cautionary note. 

\begin{figure}
\epsscale{.80}
\plotone{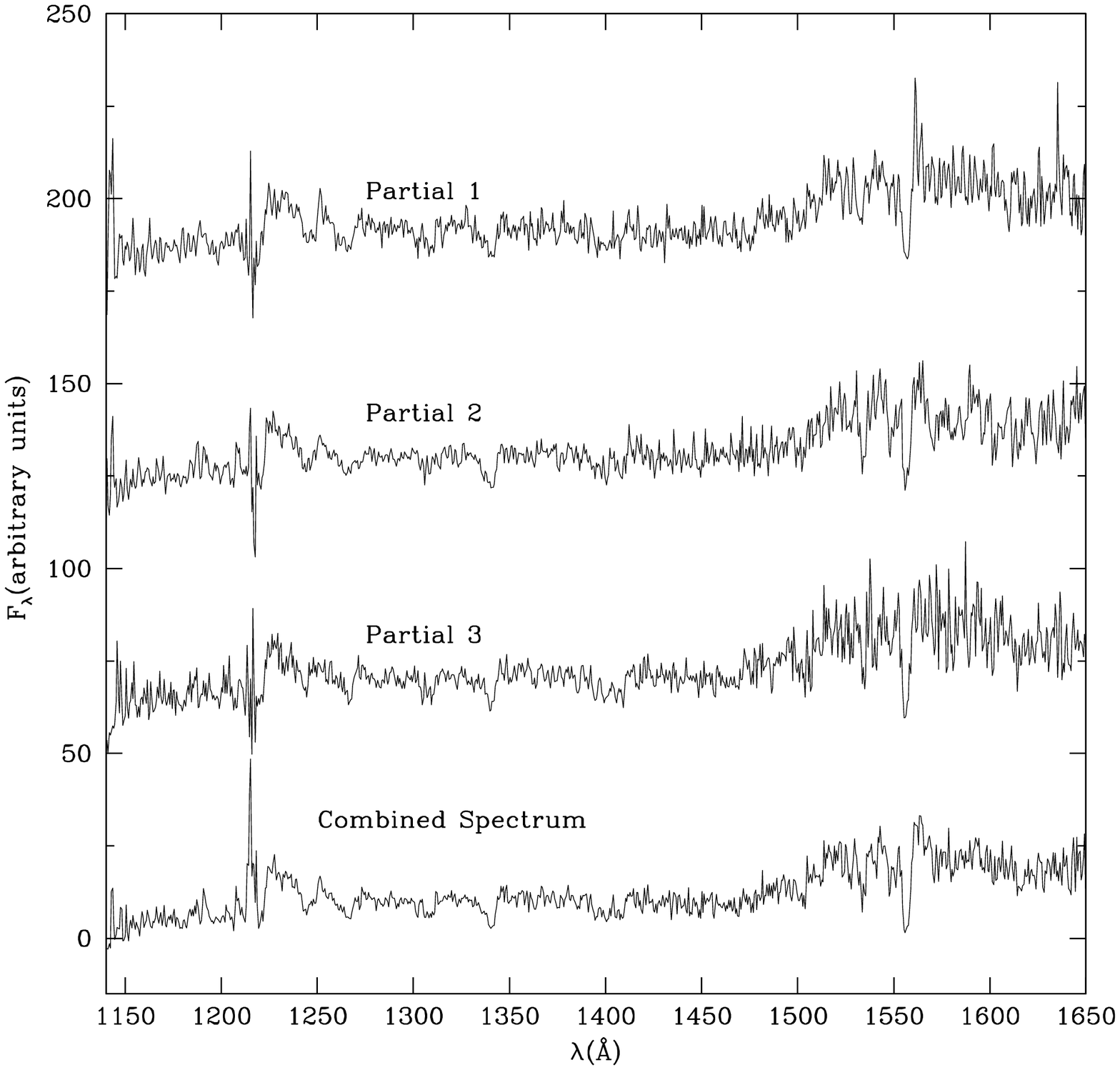}
\caption{Illustration of the process of averaging three UV spectra (observed frame) of the galaxy
NGC\,1097. 
\label{fig1}}
\end{figure}

After extracting the spectra, for the galaxies which had
more than one exposure for each spectral range (and with the same
spectral resolution, orientation and plate scale), we constructed
averages to improve the signal-to-noise ratio, eliminating also
cosmic-rays and other defects when detected. The average was only 
constructed  after checking also if the spectra had similar
flux level. For the construction of the average spectra we have used the 
task {\it scombine} in IRAF,  with the rejection algorithm 
{\it avsigclip} when three or more spectra were 
available or {\it minmax} when there were only two spectra. 
This step is illustrated in Fig. \ref{fig1}.

\section{Combination of the spectra}

\begin{figure}
\epsscale{.80}
\plotone{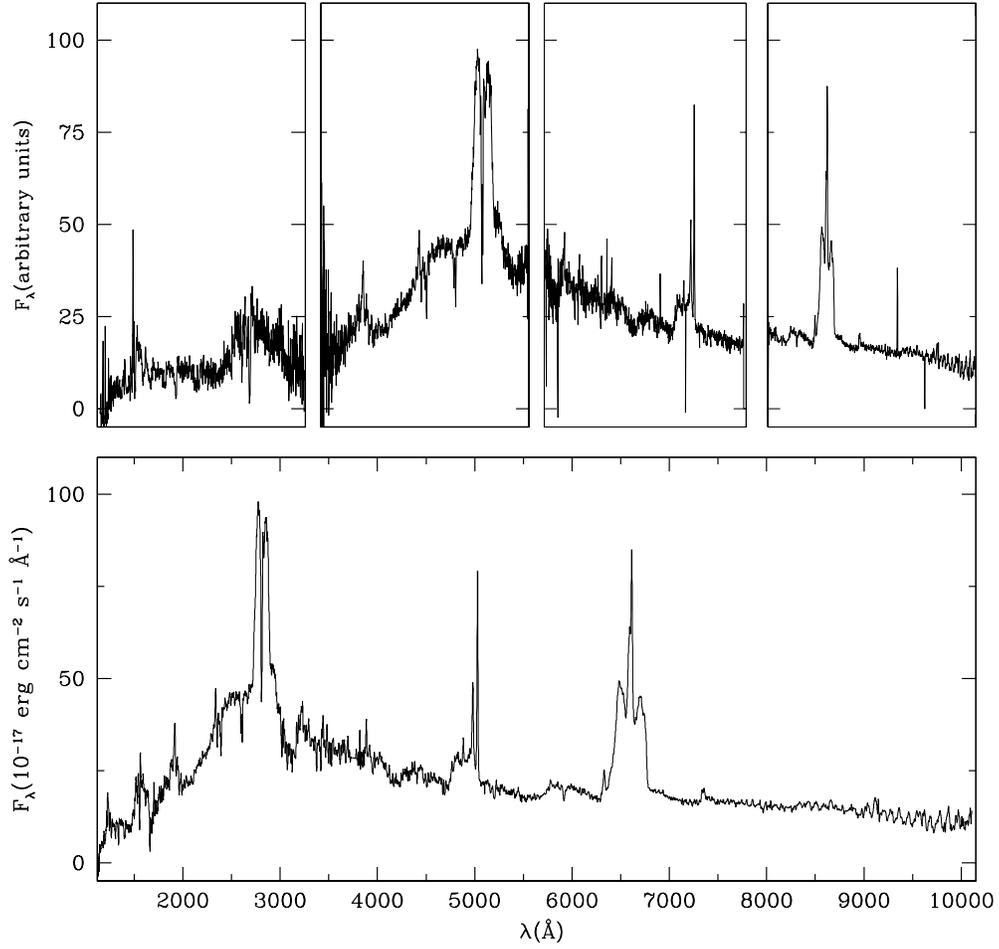}
\caption{Illustration of the process of combining different spectral segments (observed frames) for the
galaxy NGC\,1097. 
\label{fig2}}
\end{figure}

The final spectra were obtained by combining the data from the 
different spectral ranges using the same task {\it scombine} in IRAF,
after editing out noisy regions at the borders of each spectral segment,
and checking that there were no significant differences between their
fluxes. We did not find such differences
for most of the cases in which there was a significant superposition of
adjacent spectral segments. This final step is illustrated in Fig. \ref{fig2}.

In the case of the Seyfert 1 galaxies there is the issue of variability,
so that spectra obtained in different dates may show different fluxes,
in line and continua. We have checked the dates and found only 
five cases of Seyfert 1 galaxies with spectral segments obtained
in different dates: NGC\,4151, NGC\,4258, PKS\,1739+184 and AKN\,564.
In the case of IRAS\,13224-3809, 2 of the combined 
7 spectra were obtained one day latter than the other 5, thus
the effect of variability should be minimal. We have identified these
cases with a note in Table 1 and in the {\it Mastertable}. Nevertheless,
we did not find any obvious discrepancy in fluxes when combining the
different spectral segments of these galaxies.

Finally, we would like to point out that, prior to the extraction,
the flux units were $erg\,s^{-1}$\,cm$^{-2}$\,\AA$^{-1}$\,arcsec$^{-2}$ (see STIS Data Handbook).
When we performed the extraction with {\it apall}
to sum over a few pixels (0\farcs2 aperture) along the slit direction, 
the extracted spectrum is in units equivalent to $pixel \times erg\,s^{-1}$\,cm$^{-2}$\,\AA$^{-1}$\,arcsec$^{-2}$.
Then, in order to consistently provide the flux integrated in the
extraction window, in units of $erg\,s^{-1}$\,cm$^{-2}$\,\AA$^{-1}$, we multiplied each segment by a factor which is the product of the slit width and plate scale. For example, for one segment with a slit width of 0\farcs2 and platescale 0\farcs024$\,pixel^{-1}$, the factor is 
0.0048\,$arcsec^{-2}\,pixel^{-1}$. For another segment with platescale 
0\farcs05$\,pixel^{-1}$, with the same slit width, the factor is 0.01\,$arcsec^{-2}\,pixel^{-1}$.

\section{Results}

\begin{figure}
\epsscale{.80}
\plotone{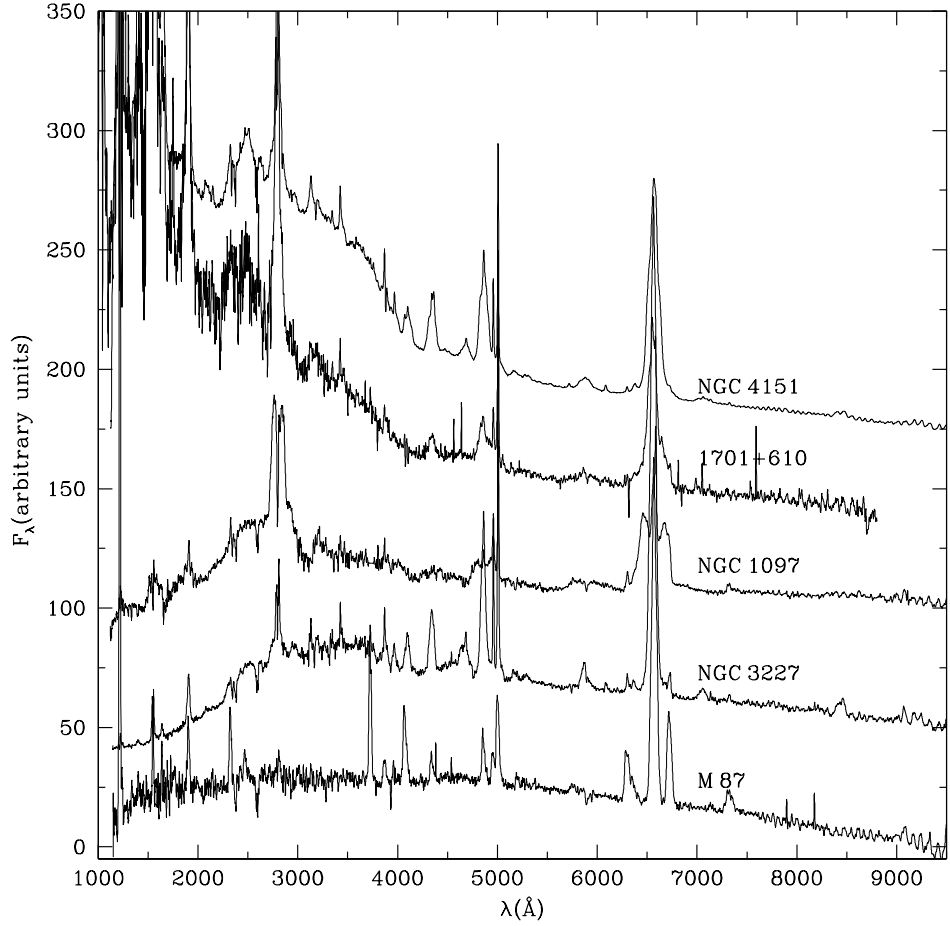}
\caption{Illustration of 5 of the 9 spectra with widest spectral coverage,
The spectra have been shifted to the rest frame of the galaxies. 
\label{fig3}}
\end{figure}

\begin{figure}
\epsscale{.80}
\plotone{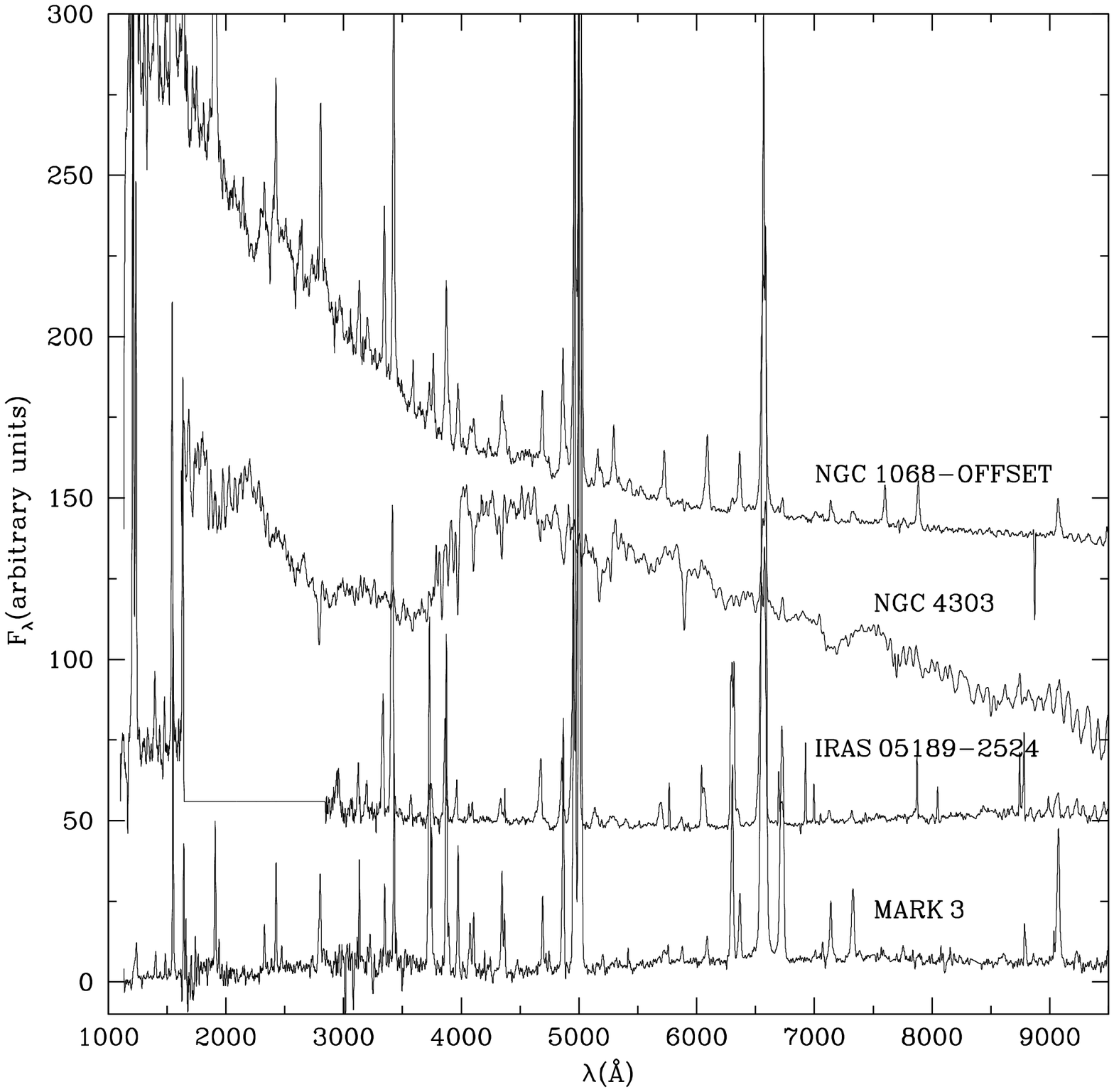}
\caption{Same as Fig. \ref{fig3} for another 4 spectra.
\label{fig4}}
\end{figure}

The combined spectra presenting the largest spectral coverage  are shown
in Figs. \ref{fig3} and \ref{fig4}. There are only 9 galaxies for 
which we could obtain the complete STIS UV-optical spectral coverage
($\sim$1000--10000\AA).

\begin{figure}
\epsscale{.80}
\plotone{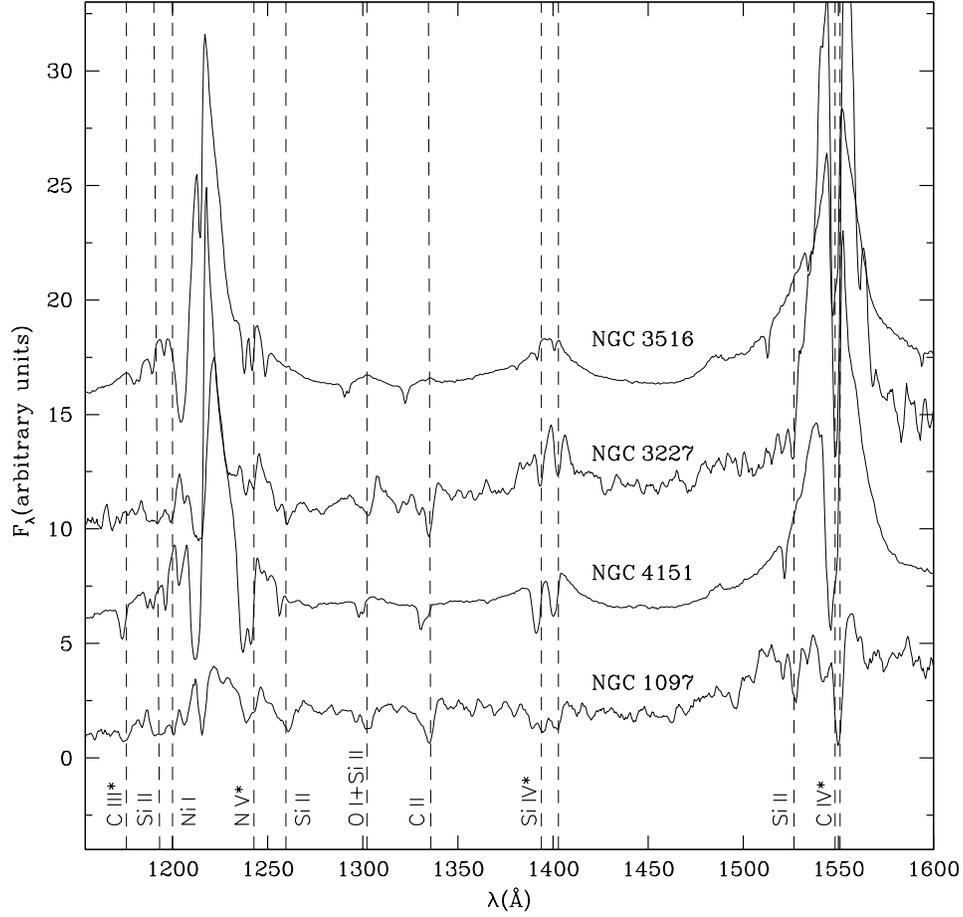}
\caption{Illustration of 4 of the 12 spectra with UV coverage in the 1100--1600\AA\, wavelength
range. The spectra have been shifted to the rest frame of the galaxies. The vertical dashed
lines show the location of absorption features typical of starbursts. Asterisks
identify the absorption lines which originate in the atmosphere of early-type stars.
\label{fig5}}
\end{figure}

\begin{figure}
\epsscale{.80}
\plotone{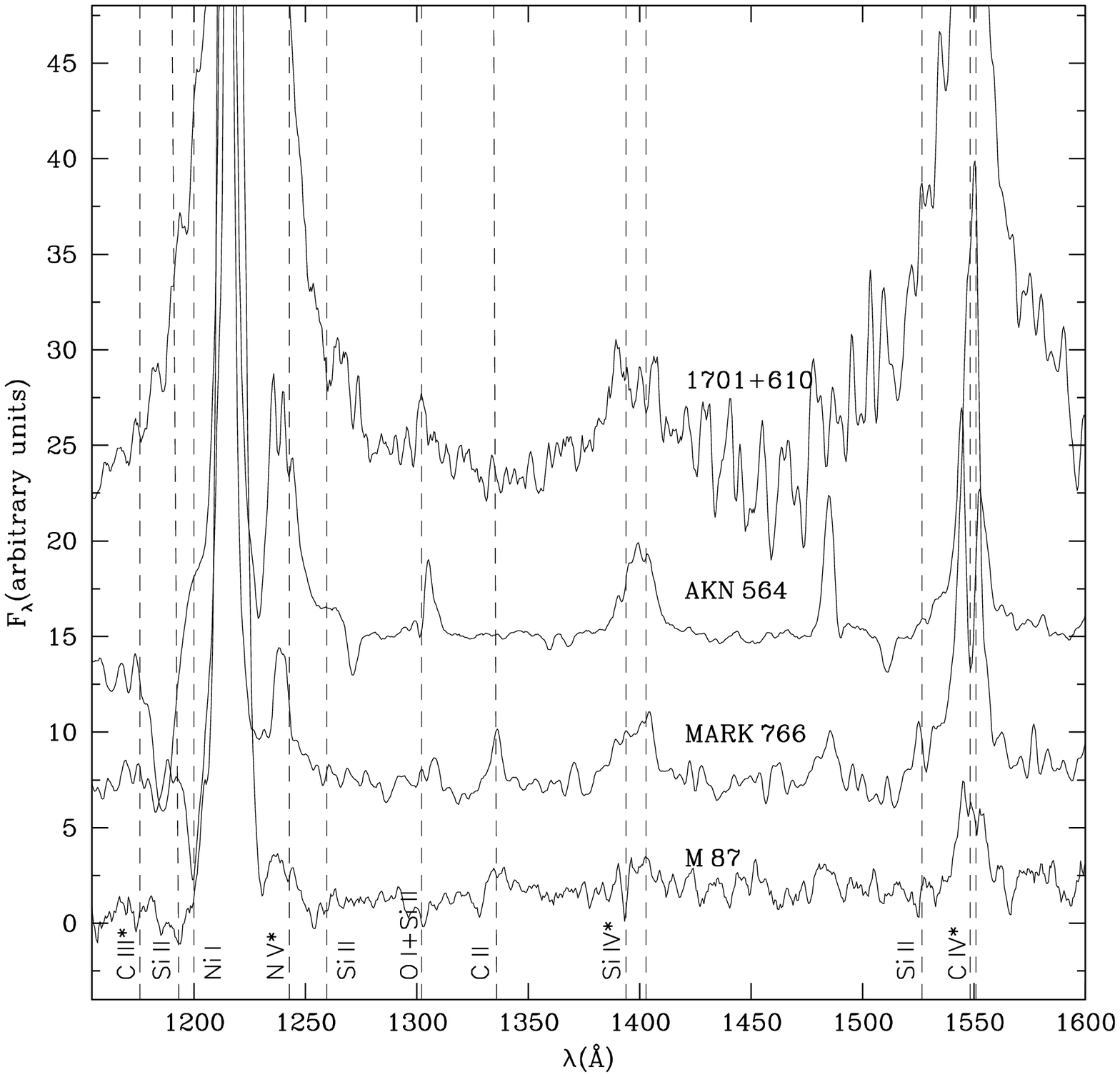}
\caption{Same as Fig. \ref{fig5} for another 4 spectra.
\label{fig6}}
\end{figure}

\begin{figure}
\epsscale{.80}
\plotone{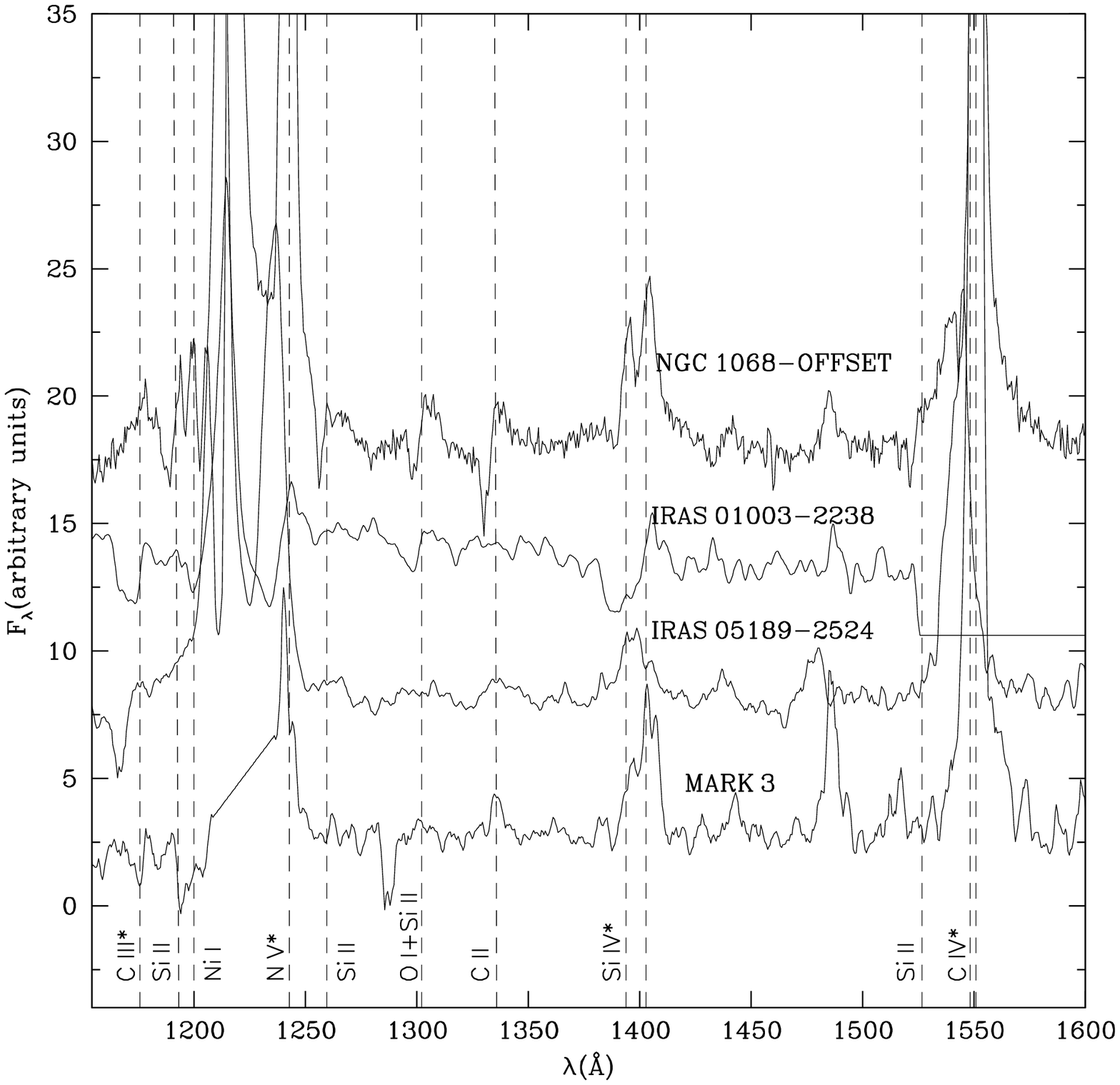}
\caption{ Same as Fig. \ref{fig5} for another 4 spectra.
\label{fig7}}
\end{figure}

In Figs. \ref{fig5}, \ref{fig6} and \ref{fig7}
we show the redshift corrected spectra for the galaxies
with UV coverage in the 1100--1600\AA\,  wavelength
range, useful for looking for signatures of starbursts. In order to do that,
we have drawn in the figures vertical lines at the locations of the absorption
features characteristic of starbursts. While most lines are interstellar,
we identify by asterisks the ones which originate in the atmosphere of
young stars \citep{kin93,vaz04}, which are 
C~III $\lambda$1175.65, N~V $\lambda\lambda$1238.81, 1242.80,
C~II $\lambda\lambda$1334.53, 1335.70, Si~IV $\lambda\lambda$1393.76, 1402.77, and C~IV $\lambda\lambda$1548.20, 1550.77. Both the interstellar and stellar 
features of a starburst have been found in the UV spectrum of 
NGC\,1097, as pointed out by \citet{sb05}. Figure \ref{fig5} shows that
the same features seem to be present in the spectrum of NGC\,3227, suggesting
that, also in this case, as in  NGC\,1097, there is a starburst
closer than 8\,pc from the nucleus in NGC\,3227 (the distance
at the galaxy corresponding to 0\farcs1 the angular distance
from the nucleus covered by the aperture of the nuclear extraction).
Indeed, the presence of traces of young stellar population
in an optical nuclear spectrum of NGC\,3227 have
been previously reported by \citet{gon97}.
These features seem also to be present in the spectrum of NGC\,4151,
but this has to be investigated further, as they may be due 
to absorptions in our galaxy, due to the proximity of NGC\,4151.
An obvious case of interstellar absorptions from our galaxy can
be observed in the UV spectrum of NGC\,3516 (Fig. \ref{fig5}),
where absorptions from O\,I\,$\lambda$1302.08\,+\,Si\,II\,$\lambda$1304.40 
and C\,II\,$\lambda\lambda$1334.53,1335.70 originating in the
Milky Way appear blueshifted from their wavelengths due to the shift of the spectrum to the rest frame of the galaxy.

All spectra are available electronically at the URL address:
{\bf www.if.ufrgs.br/\~{}pat/atlas.htm}, where they can
be visualized and recovered by clicking on the name of the galaxy.
We also make available at the above address the {\it Mastertable},
which has a compilation of the relevant information on each
spectrum used in the combination.

Finally, we point out that there are several spectra which were obtained 
only with the highest resolution gratings, therefore covering
a short wavelength range. In many cases there is also a sequence of such spectra obtained
at adjacent slit positions, apparently for kinematic studies. In these cases,
we did not combine the spectra but provide instead the individual
extractions in a {\it tar} file.

\clearpage

\clearpage

\clearpage

\clearpage

\clearpage

\clearpage

\clearpage

\clearpage

\begin{deluxetable}{llllllll}
\tabletypesize{\scriptsize}
\tablecaption{Galaxy sample\label{tbl-1}}
\tablewidth{0pt}
\tablehead{
\colhead{Galaxy} & \colhead{RA\tablenotemark{a}} & \colhead{DEC\tablenotemark{a}} & \colhead{Hubble Type\tablenotemark{a}} & \colhead{Activity\tablenotemark{a}} &
\colhead{Z\tablenotemark{a}} & \colhead{Coverage} & \colhead{References} \\
\colhead{} & \colhead{(hms)} & \colhead{($^\circ$ $^{\prime}$ $^{\prime\prime}$)} & \colhead{} & \colhead{} &
\colhead{} & \colhead{(\AA)} & \colhead{}}
\startdata
Q0038+327\tablenotemark{b} &00 40 43.5 &+32 58 33 &-- &Sy? &0.1970 &1640-3175\tablenotemark{b} &--\\
MARK348	&00 48 47.1 &+31 57 25 &-- &HII/WR, Sbrst, Sy2 &0.1177 &2500-5700 &-\\
IRAS01003-2238 &01 02 49.9 &-22 21 56 &SB(rs)bc &Sy &0.0049 &1140-10226 &1\\
NGC613 &01 34 18.2 &-29 25 07 &SA(s)0/a &Sy2 &0.0150 &6482-7054 &2,3\\
MARK573 &01 43 57.8 &+02 21 00 &(R)SAB(rs)0+ &Sy2 &0.0172 &2900-6867 &4\\
UM146 &01 55 22.0 &+06 36 43 &SA(rs)b &Sy1.9 &0.0174 &2900-6867 &4 \\
NGC788 &02 01 06.4 &-06 48 56 &SA(s)0/a &Sy1, Sy2 &0.0136 &2900-6867 &4\\
3C67 &02 24 12.3 &+27 50 12 &-- &BLRG &0.3102 &2900-10226 &5\\
NGC985 &02 34 37.8 &-08 47 15 &SBbc?p(Ring) &Sy1 &0.0431 &1194-1250 &6\\
NGC1052 &02 41 04.8 &-08 15 21 &E4 &LINER, Sy2	&0.0049 &6295-6867 &7,8 \\
NGC1068	&02 42 40.7 &-00 00 48 &(R)SA(rs)b &Sy1, Sy2 &0.0038 &1140-10266 &9,10\\
NGC1097\tablenotemark{d}&02 46 19.0 &-30 16 30 &(R$^{\prime}$\_1:)SB(r$^{\prime}$l)b &Sy1 &0.0042 &1140-10266\tablenotemark{d} &--\\
MARK1066 &02 59 58.6 &+36 49 14 &(R)SB(s)0+ &Sy2 &0.0120 &2900-5700 &-- \\
NGC1358 &03 33 39.7 &-05 05 22 &SAB(r)0/a &Sy2	&0.0134 &2900-6867 &4 \\
MS0335.4-2618\tablenotemark{b} &03 37 36.6 &-26 09 08 &-- &Sy1 &0.1230 &1150-1740\tablenotemark{b} &--\\
3C109 &04 13 40.4 &+11 12 14 &Opt.var &Ngal, Sy1.8 &0.3056 &2900-10266 &--\\
3C120 &04 33 11.1 &+05 21 16 &S0,LPQ & BLRG, Sy1 &0.0330 &2900-10266 &--\\
MARK618\tablenotemark{b}&04 36 22.2 &-10 22 34 &SB(s)b pec &Sy1 &0.0355 &1640-3175\tablenotemark{b} &11\\
NGC1667 &04 48 37.1 &-06 19 12 &SAB(r)c	&Sy2 &0.0152 &2900-6867 &4\\
3C135 &05 14 08.3 &+00 56 32 &E &BLRG, Sy2 &0.1274 &5236-10266 &12\\
AKN120\tablenotemark{b}&05 16 11.4 &-00 08 59 &Sb/pec &Sy1 &0.0323 &1640-3175\tablenotemark{b} &11\\
IRAS05189-2524 &05 21 01.3 &-25 21 45 &pec &Sy2 &0.0426 &1140-10266 &1\\
NGC1961	&05 42 04.8 &+69 22 43 &SAB(rs)c &LINER &0.0131 &6295-6867 &--\\
NGC2110 &05 52 11.4 &-07 27 22 &SAB0- &Sy2 &0.0078 &6295-6867 &13\\
MARK3 &06 15 36.3 &+71 02 15 &S0 &Sy2 &0.0135 &1140-10266 &14,15\\
NGC2273 &06 50 08.7 &+60 50 45 &SB(r)a &Sy2 &0.0062 &2900-6867 &4\\
MARK9\tablenotemark{b} &07 36 57.0 &+58 46 13 &S0 pec? &Sy1.5 &0.0399 & 1640-3175\tablenotemark{b} &11\\
MARK78\tablenotemark{b} &07 42 41.7 &+65 10 37 &SB &Sy2 &0.0371 & 1140-7054\tablenotemark{b} &16\\
NGC2787 &09 19 18.5 &+69 12 12 &SB(r)0+	&LINER &0.0023 &2900-6867 &17,18,19,20\\
NGC2841	&09 22 02.6 &+50 58 35 &SA(r)b	&LINER, Sy1 &0.0021 &8275-8847 &--\\
MARK110	&09 25 12.9 &+52 17 11 &Pair? &Sy1 &0.033 &1194-1250 &--\\
NGC2911	&09 33 46.1 &+10 09 09 &SA(s)0:pec &LINER, Sy &0.0106 &6482-7054 &--\\
NGC3031	&09 55 33.2 &+69 03 55 &SA(s)ab	&LINER, Sy1.8 &-0.0001 &8275-8847/6265-6867 &21,22\\
NGC3081	&09 59 29.5 &-22 49 35 &(R\_1)SAB(r)0/a &Sy2 &0.0079 &2900-6867 &4\\
MARK34 &10 34 08.6 &+60 01 52 &Spiral &Sy2 &0.0505 &2900-5700 &--\\
NGC3227	&10 23 30.6 &+19 51 54 &SAB(s)pec &Sy1.5 &0.0039 &1140-10266 &7,8,23\\
NGC3393\tablenotemark{d}&10 48 23.4 &-25 09 43 &(R$^{\prime}$)SB(s)ab &Sy2 &0.0125 &2900-6867\tablenotemark{d} &24\\
NGC3516	&11 06 47.5 &+72 34 07 &(R)SB(s)0\^{}0\^{} &Sy1.5 &0.0088 &1140-5700/6265-6867 &24,25\\
IRAS11058-1131 &11 08 20.3 &-11 48 12 &-- &Sy2 &0.0548 &2900-6867 &24\\
ESO438-G009\tablenotemark{d} &11 10 48.0 &-28 30 04 &(R$^{\prime}$\_1)SB(rl)ab &Sy1.5 &0.0234 &1194-1250\tablenotemark{d} &26\\
MCG10.16.111 &11 18 57.7 &+58 03 24 &-- &Sy1 &0.0279 &1194-1250  &26\\
NGC3627	&11 20 15.0 &+12 59 30 &SAB(s)b	&LINER, Sy2 &0.0024 &2900-6867 &--\\
SBS1127+575\tablenotemark{b} &11 30 03.6 &+57 18 29 &-- &Sy2 &0.0361 & 1194-1250\tablenotemark{b} &26\\
PG1149-110 &11 52 03.5 &-11 22 24 &-- &Sy1 &0.0490 &1194-1250 &26\\
NGC3982	&11 56 28.1 &+55 07 31 &SAB(r)b &Sy2 &0.0037 &2900-6867 &17,18,19,20\\
NGC3998	&11 57 56.1 &+55 27 13 &SA(r)0\^{}0\^{}? &LINER, Sy1 &0.0035 &8275-8847 &--\\
NGC4036	&12 01 26.9 &+61 53 44 &S0- &LINER &0.0048 &6295-6867 &7,8\\
3C268.3	&12 06 24.7 &+64 13 37 &-- &BLRG &0.3710 &5236-10266 &12\\
NGC4138	&12 09 29.6 &+43 41 17 &SA(r)0+ &Sy1.9 &0.0030 &2900-6867 &17,18,19,20\\
IRAS12071-0444 &12 09 45.1 &-05 01 14 &-- &Sy2 &0.1283 &5236-10266 &1\\
NGC4151\tablenotemark{d} &12 10 32.6 &+39 24 21 &(R$^{\prime}$)SAB(rs)ab &Sy1.5 &0.0033 &1140-10266\tablenotemark{d} &27 to 33\\
MARK766	&12 18 26.5 &+29 48 46 &(R$^{\prime}$)SB(s)a &Sy1.5 &0.0129 &1140-3184 &--\\
NGC4258\tablenotemark{d} &12 18 57.5 &+47 18 14 &SAB(s)bc &LINER, Sy1.9 &0.0015 &8275-8847\tablenotemark{d} &2,3\\
NGC4278	&12 20 06.8 &+29 16 51 &E1-2 &LINER, Sy1 &0.0022 &8275-8847 &--\\
Q1219+047\tablenotemark{b} &12 21 37.9 &+04 30 26 &-- &Sy1 &0.0940 &1194-1250\tablenotemark{b} &--\\
NGC4303	&12 21 54.9 &+04 28 25 &SAB(rs)bc &HIISy2 &0.0052 &1568-10266 &2,3,34\\
NGC4450	&12 28 29.6 &+17 05 06 &SA(s)ab	&LINER, Sy3 &0.0065 &2900-10266 &17,18,19,20\\
NGC4477	&12 30 02.2 &+13 38 11 &SB(s)0? &Sy2 &0.0045 &2900-6867 &17,18,19,20\\
M87 &12 30 49.4 &+12 23 28 &E+0-1pec &NLRG, Sy &0.0044 &1140-10266 &35\\
NGC4501	&12 31 59.2 &+14 25 14 &SA(rs)b &Sy2 &0.0076 &2900-6867 &17,18,19,20\\
TON1542	&12 32 03.6 &+20 09 29 &Spiral &Sy1 &0.0630 &1194-1300 &6\\
NGC4540\tablenotemark{b}&12 34 50.8 &+15 33 05 &SAB(rs)cd &LINER, Sy1 &0.0043 &2900-5700\tablenotemark{b} &--\\
NGC4507	&12 35 36.6 &-39 54 33 &SAB(s)ab &Sy2 &0.0118 &2900-6867 &4\\
NGC4569	&12 36 49.8 &+13 09 46 &SAB(rs)ab &LINER, Sy &-0.0008 &2900-6867 &--\\
NGC4579	&12 37 43.6 &+11 49 05 &SAB(rs)b &LINER, Sy1.9 &0.0051 &6295-6867 &7,8\\
NGC4594	&12 39 59.4 &-11 37 23 &SA(s)a &LINER, Sy1 &0.0034 &6482-7054 &--\\
IC3639 &12 40 52.8 &-36 45 21 &SB(rs)bc &Sy2 &0.0109 &2900-6867 &4\\
NGC4698	&12 48 22.9 &+08 29 14 &SA(s)ab &Sy2 &0.0033 &2900-6867 &17,18,18,20\\
NGC4736	&12 50 53.0 &+41 07 14 &(R)SA(r)ab &LINER, Sy2 &0.0010 &6295-6867 &--\\
NGC4826	&12 56 43.7 &+21 40 52 &(R)SA(rs)ab &Sy2 &0.0014 &2900-6867 &--\\
NGC5005	&13 10 56.2 &+37 03 33 & SAB(rs)bc &Sy2,LINER &0.0032 &6482-7054 &2,3\\
IRAS13224-3809\tablenotemark{d} &13 25 19.3 &-38 24 53 &-- &Sbrst, NLSy1 &0.0667 &5236-10266\tablenotemark{d} &36,37\\
NGC5135	&13 25 44.0 &-29 50 01 &SB(l)ab	&Sy2 &0.0137 &2900-5700/6295-6768 &4\\
NGC5194\tablenotemark{c} &13 29 52.7 &+47 11 43 &SA(s)bc pec &HIISy2.5 &0.0015 & 2900-10266\tablenotemark{c} &38\\
NGC5252	&13 38 15.9 &+04 32 33 &S0 &Sy1.9 &0.0230 &2900-5700 &24\\
NGC5283 &13 41 05.7 &+67 40 20 &S0? &Sy2 &0.0104 &2900-6867 &4\\
TON730 &13 43 56.7 &+25 38 48 &-- &Sy1 &0.0870 &1194-1250 &26\\
NGC5347	&13 53 17.8 &+33 29 27 &(R$^{\prime}$)SB(rs)ab &Sy2 &0.0078 &2900-6867 &4\\
MARK463E &13 56 02.9 &+18 22 19 &-- &Sy1, Sy2 &0.0500 &2900-5700 &-\\
NGC5427	&14 03 26.0 &-06 01 51 &SA(s)c, pec &Sy2 &0.0087 &2900-6867 &4\\
Circinus &14 13 09.9 &-65 20 21 &SA(s)b &Sy2 &0.0014 &4818-5104 &-\\
NGC5635	&14 28 31.7 &+27 24 32 &S, pec &LINER, Sy3 &0.0144 &6482-7054 &-\\
NGC5643	&14 32 40.8 &-44 10 29 &SAB(rs)c &Sy2 &0.0040 &2900-6867 &4\\
MARK817\tablenotemark{b} &14 36 22.1 &+58 47 39 &SBc &Sy1.5 &0.0314 & 2758-2914\tablenotemark{b} & 39\\
NGC5695	&14 37 22.1 &+36 34 04 &SBb &Sy2 &0.0141 &2900-6867 &4\\
NGC5728 &14 42 23.9 &-17 15 11 &(R\_1)SAB(r)a &Sy2 &0.0093 &6295-6867 &-\\
IRAS15206+3342\tablenotemark{b} &15 22 38.0 &+33 31 36 & ? &HIISy2 &0.1244 & 1140-10266\tablenotemark{b} &1\\
3C346 &16 43 48.6 &+17 15 49 &E &NLRG, Sy2 &0.1620 &2900-10266 &12\\
1701+610 &17 02 11.1 &+60 58 48 &-- &Sy1.9 &0.1649 &1140-10266 &-\\
NGC6300	&17 16 59.5 &-62 49 14 &SB(rs)b &Sy2 &0.0037 &6581-6867 &4\\
PKS1739+184\tablenotemark{d} &17 42 06.9 &+18 27 21 &-- &Sy1 &0.1860 &1140-5700\tablenotemark{d} &-\\
3C405\tablenotemark{b}&19 59 28.3 &+40 44 02 &S? &Radiogal, Sy2 &0.0561 &2900-5700\tablenotemark{b} &40,41\\
3C382 &18 35 02.1 &+32 41 50 &-- &BLRG, Sy1 &0.0579 &2900-10266 &-\\
3C390 &18 45 37.6 &+09 53 45 &-- &RadioS	&-- &2900-5700 &-\\
NGC6951 &20 37 14.1 &+66 06 20 &E+pec? &-- &0.0129 &6482-7054 &2,3\\
3C445 &22 23 49.6 &-02 06 12 &N galaxy &BLRG, Sy1 &0.0562 &2900-10266 &-\\
NGC7314 &22 35 46.2 &-26 03 01 &SAB(rs)bc &Sy1.9 &0.0048 &2900-10266 &2,3\\
AKN564\tablenotemark{d} &22 42 39.3 &+29 43 31 &SB &Sy1.8 &0.0247 &1140-3184\tablenotemark{d} &43,44,45\\
IC1459 &22 57 10.6 &-36 27 44 &E3 &LINER &0.0056 &2900-5700 &42\\
NGC7674	&23 27 56.7 &+08 46 45 &SA(r)bc pec &HIISy2 &0.0289 &2900-5700 &-\\
NGC7682 &23 29 03.9 &+03 32 00 &SA(r)bc pec &HIISy2 &0.0289 &2900-6867 &4\\

\enddata
\tablenotetext{a}{References from NASA/IPAC Extragalactic Database.}
\tablenotetext{b}{Spectra of this galaxy were not extracted
due to a poor signal-to-noise ratio in the continuum. Nevertheless, information on the available
spectra is also included in the {\it Mastertable}.}
\tablenotetext{c}{Spectra of this galaxy were not extracted
due to the presence of more than one continuum source which we could not identify the brightest one. Information on the available
spectra is also included in the {\it Mastertable}.}
\tablenotetext{d}{Final spectrum of this Seyfert 1 galaxy was obtained with spectra observed in different dates.}

\tablecomments{References:1-\citet{far05}, 2-\citet{hu03}, 3-\citet{hu05}, 4-\citet{po03}, 5-\citet{odea03}, 6-\citet{pen04}, 7-\citet{ba01a}, 8-\citet{ba01b}, 9-\citet{kra00b}, 10-\citet{ce02}, 11-\citet{je03}, 12-\citet{hut98}, 13-\citet{fe04}, 14-\citet{colls05}, 15-\citet{ru01}, 16-\citet{whi05}, 17-\citet{sar01}, 18-\citet{sar02}, 19-\citet{sar05}, 20-\citet{ho02}, 21-\citet{cha01a}, 22-\citet{cha01b}, 23-\citet{cre01}, 24-\citet{capt05}, 25-\citet{ed00}, 26-\citet{bo02}, 27-\citet{kai00}, 28-\citet{kra00a}, 29- \citet{nel00}, 30-\citet{hut99}, 31-\citet{cre00}, 32-\citet{hut02}, 33-\citet{kra01}, 34-\citet{col02}, 35-\citet{sab03}, 36-\citet{lei04a}, 37-\citet{lei04b}, 38-\citet{bra04}, 39-\citet{je03}, 40-\citet{tad03}, 41-\citet{bel04}, 42-\citet{cap02}, 43-\citet{cre02}, 44-\citet{collr01} and 45-\citet{rom04}.}
\end{deluxetable}
\clearpage

\begin{deluxetable}{lccccccccccc}
\tabletypesize{\scriptsize}
\rotate
\tablecaption{A sample of lines from the {\it Mastertable}\label{tbl-2}}
\tablewidth{0pt}
\tablehead{
\colhead{(1)} & \colhead{(2)} & \colhead{(3)} & \colhead{(4)} & \colhead{(5)} & \colhead{(6)} & \colhead{(7)} & \colhead{(8)} & \colhead{(9)} & \colhead{(10)} &
\colhead{(11)} & \colhead{(12)}\\
\colhead{Galaxy} & \colhead{Rootname} & \colhead{Grating} & \colhead{Aperture} & \colhead{ $\lambda_c$} & \colhead{$\lambda_i$} & \colhead{$\lambda_f$} & \colhead{R} & \colhead{PA} & \colhead{Exp. Time} &
\colhead{Name} & \colhead{Scale} \\
\colhead{} & \colhead{} & \colhead{} & \colhead{($^{\prime\prime}$ $^{2}$)} & \colhead{(\AA)} & \colhead{(\AA)} & \colhead{(\AA)} & \colhead{} & \colhead{($^\circ$)} & \colhead{(s)} &
\colhead{} & \colhead{($^{\prime\prime}$ $pix^{-1}$)}}

\startdata
\tableline
MCG10.16.111 &o5ew02010 &G140M &52x0.2 &1222 &1194 &1250 &12200 &-86.5063 &3900 &m1016111-1.234 &0.029\\
 & o5ew02020 &G140M &52x0.2 &1222 &1194 &1250 &12200 &-86.5063 &3900 &m1016111-2.234 &0.029\\
 & o5ew02030 &G140M &52x0.2 &1222 &1194 &1250 &12200 &-86.5063 &3900 &m1016111-3.234 &0.029\\
 & o5ew02040 &G140M &52x0.2 &1222 &1194 &1250 &12200 &-86.5063 &3900 &m1016111-4.234 &0.029\\
 & o5ew02050 &G140M &52x0.2 &1222 &1194 &1250 &12200 &-86.5062 &3900 &m1016111-5.234 &0.029\\

\tableline
NGC3627	&o63n02010 &G430L &52x0.2 &4300 &2900 &5700 &800 &80.0559 &2349 &n3627-1.80 &0.1\\
 &o63n02020 &G750M &52x0.2 &6581 &6295 &6867 &5980 &80.0559 &2861 &n3627-2.80 &0.1\\

\tableline
PG1149-110 &o5ew05010 &G140M &52x0.2 &1222 &1194 &1250 &12200 &43.6861 &2269 &p1149-1.44 &0.029\\
 &o5ew05020 &G140M &52x0.2 &1222 &1194 &1250 &12200 &43.6861 &2899 &p1149-2.44 &0.029\\
 &o5ew05030 &G140M &52x0.2 &1222 &1194 &1250 &12200 &43.6861 &2899 &p1149-3.44 &0.029\\

\tableline
NGC3982	&o4e006010 &G750M &52X0.2 &6581 &6295 &6867  &5980 &117.931 &900 &n3982-1.117 &0.05\\
 &o4e006020 &G750M &52X0.2 &6581 &6295 &6867  &5980 &117.931 &1197 &n3982-2.117 &0.05\\
 &o4e006030 &G750M &52X0.2 &6581 &6295 &6867  &5980 &117.931 &900 &n3982-3.117 &0.05\\
 &o4e006040 &G430L &52X0.2 &4300 &2900 &5700  &800  &117.931 &900 &n3982-4.117 &0.05\\
 &o4e006050 &G430L &52X0.2 &4300 &2900 &5700  &800  &117.931 &945 &n3982-5.117 &0.05\\

\tableline
IRAS15206+3342 &o5f904030 &G430L &52x0.2 &4300 &2900 &5700 &800 &35.3559 &780 &i1520-3.35 &0.05\\
 &o5f904040 &G430L &52x0.2 &4300 &2900 &5700 &800 &35.3559 &650 &i1520-4.35 &0.05\\
 &o5f904050 &G750L &52x0.2 &7751 &5236 &10266 &790 &35.356 &624 &i1520-5.35 &0.05\\
 &o5f904060 &G750L &52x0.2 &7751 &5236 &10266 &790 &35.3559 &624 &i1520-6.35 &0.05\\
 &o5f904070 &G750L &52x0.2 &7751 &5236 &10266 &790 &35.3559 &545 &i1520-7.35 &0.05\\
 &o5f904090 &G140L &52x0.2 &1425 &1140 &1730 &1190 &35.3001 &900 &i1520-8.35 &0.0244\\
\enddata
\tablecomments{Columns: (1) the name of the galaxy; (2) the identification of all 
available STIS spectra for this galaxy in the HST archive,
one per line; (3) the grating used in each observation; (4) the slit width of each observation; 
(5) the central wavelength; (6) the initial wavelength; (7) the final wavelength;
(8) the spectral resolution; (9) the slit orientation; (10) the exposure time; 
(11) The identification of the extracted spectrum from each
segment; (12) platescale of the observations.}
\end{deluxetable}
\clearpage

\end{document}